# Optothermal needle-free injection of vaterite nanocapsules

Denis Kislov[1,=], Daniel Ofer[2,3,=], Andrey Machnev,[2,3,=] Hani Barhom[2,3,4], Vjaceslavs Bobrovs[5], Alexander Shalin[1,6], and Pavel Ginzburg[2,3] *

[1] Center for Photonics and 2D Materials, Moscow Institute of Physics and Technology, Dolgoprudny, 141700, Russia
[2] Department of Electrical Engineering, Tel Aviv University, Ramat Aviv, Tel Aviv 69978, Israel
[3] Light-Matter Interaction Centre, Tel Aviv University, Tel Aviv, 69978, Israel
[4] Triangle Regional Research and Development Center, Kfar Qara' 3007500, Israel
[5] Riga Technical University, Institute of Telecommunications, Riga 1048, Latvia
[6] Faculty of Physics, M. V. Lomonosov Moscow State University, 119991 Moscow, Russia

**Abstract**

The propulsion and acceleration of nanoparticles with light have both fundamental and applied significance across many disciplines. Needle-free injection of biomedical nano cargoes into living tissues is among the examples. Here we explore a new physical mechanism of laser-induced particle acceleration, based on abnormal optothermal expansion of mesoporous vaterite cargoes. Vaterite nanoparticles, a metastable form of calcium carbonate, were placed on a substrate, underneath a target phantom, and accelerated towards it with the aid of a short femtosecond laser pulse. Light absorption followed by picosecond-scale thermal expansion was shown to elevate the particle's center of mass thus causing acceleration. It was shown that a 2μm size vaterite particle, being illuminated with 0.5 W average power 100 fsec IR laser, is capable to overcome van der Waals attraction and acquire 15m/sec velocity. The demonstrated optothermal laser-driven needle-free injection into a phantom layer promotes the further development of light-responsive nanocapsules, which can be equipped with additional optical and biomedical functions for delivery, monitoring, and controllable biomedical dosage to name a few.

*corresponding author: denis.a.kislov@gmail.com

# 1. Introduction

Needle-free injection of biomedical cargoes into living tissues allows for reducing possible contamination and suppresses inflating unnecessary pain in patients. While quite a few different proposals including laser-based solutions have been reported [1], [2], controllable injection of a single biomedical nanocargo remains a challenge. Here we will explore the capabilities of light-driven tools to address this challenge. Since the first demonstration by Ashkin in 1970 [3], optomechanical tools were significantly advanced and taken towards practical applications [4], where propulsion, acceleration, and trapping of nanoparticles with light are of use [5]. Optical tweezers have been found beneficial in biological studies, as they allow controlling the motion of micro- and nano-scale objects, unfolding proteins and molecules, measuring pico-Newton scale forces, and can grant many other capabilities [6]–[8].

Optical forces, acting on isolated systems, are quite well understood and have been assessed in many experiments. In terms of electromagnetic theory, Maxwell's stress tensor linking classical light-matter interactions with Newtonian forces provides a comprehensive description of the phenomenon [9]. The formalism can be further extended to cavity quantum optomechanical regimes [10], where many remarkable effects can be observed, though demanding isolation from an environment, e.g., by cryogenic cooling and high vacuum [11], [12]. In room-temperature biological scenarios, however, a range of different physical mechanisms can contribute to interactions. In many cases, thermal forces, emerging from environmental interactions, can prevail over optical and, in fact, govern the dynamics. One of the key contributing effects affecting the optical manipulation of nano- and micro-objects is thermal noise [13], [14]. Fluctuations affect particle-environment interactions [15], [16], imposing additional constraints on optical tweezing in biological media [17]. Furthermore, temperature gradients in an embedding fluid give rise to directional thermophoretic forces acting on small particles. Apart from external non-isothermal conditions, those gradients can be locally induced by light absorption [18], [19]. The optically-controlled thermophoretic effect can be utilized for propelling nano- and micro-objects in fluids and can predominate over purely electromagnetic optomechanical forces [20], [21]. For example, local laser cooling of a substrate led to colloidal particles and molecules trapping in a low-temperature region - the so-called - opto-refrigerative tweezers [22]. Colloidal objects can also be heated directly with a laser, thus inspiring self-thermophoresis. Nonuniform heating of an absorbing particle, caused by a nonsymmetric illumination, inspires strong self-thermophoretic forces, e.g. [23], [24]. The symmetry in heating can be broken by an object's design. So-called Janus particles are typically realized as transparent glass spheres, half-coated with strongly absorbing gold [25]–[27]. Optothermal manipulation schemes were further explored for capturing and separating particles [28]–[30]. Relying on the above, understanding optothermal effects is critical to developing advanced tools for tailoring the mechanical motion of nano- and micro-objects iterating with an environment.

An appealing approach to investigate temperature effects in optomechanical scenarios is to compare trapping with monochromatic (continuous wave, CW) and pulsed (e.g., femtosecond) lasers, both having the same average power. Several experimental reports, e.g., [31], [32], explored this scenario, and no qualitative differences in optomechanical behavior have been found, apart from strong optical nonlinearities, inspired by high peak powers of femtosecond sources [33]–[35]. Since those experiments have been performed in a liquid environment, overdamped mechanical motion and heat diffusion average the impact of short pulses over a repetition rate of a source. To factor out the fluid effects, trapping can be performed in a vacuum, e.g., [36], [37]. However, the studies rarely consider mechanical deformations of particles themselves, at least as being the major effect, which governs the interaction.

Here we demonstrate a new optomechanical effect, exploring the interaction with short intense laser pulses with particles on a substrate (Fig. 1). Observation of the effect requires meeting several conditions on the particle and optical arrangement, as it will become evident hereinafter. The practical outlook on the effect is the needle-free injection, where a drug cargo is accelerated with light towards tissue and penetrates it.

The manuscript is organized as follows. First, the basic experiment on particle acceleration is demonstrated and release velocities are assessed. After the effect demonstration, basic physical mechanisms, which might contribute to the interaction are discussed. To figure out which one is indeed the most contributing, an additional set of experimental investigations is performed. Finally, particle penetration capabilities are assessed on pathways toward a new needle-free injection methodology.

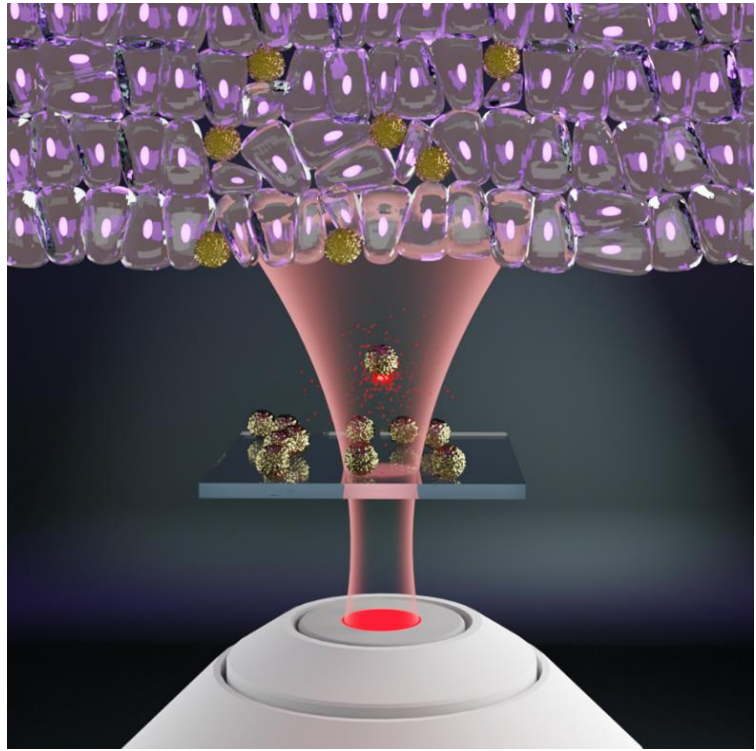

**Fig.1.** The concept of needle-free nanoparticle injection - nanocargoes on a substrate are accelerated with a short femtosecond laser towards a target tissue. Optothermal expansion is the underlying mechanism, which governs the light-particle interaction.

2. **Observation of the particle's jump**

The experiment was performed with 2μm radius vaterite nanoparticles, Fig. 2(b). Vaterite is a metastable polymorph of calcium carbonate ($CaCO_3$ [38], [39]). Owing to its high load capacity, chemical non-specific loading, biodegradability, biocompatibility [40], and facile low-cost fabrication, vaterite nanoparticles are an extremely promising nonorganic platform for drug delivery applications [41]–[43]. Furthermore, those cargoes can be designed in different shapes, which affects their optical and biological functions [44]. Mesoporous particles can be also loaded with contrast inclusions, e.g., metal nanoseeds [45], which severely affects the light-matter interaction strength of the composite, as an outlook on future optomechanical manipulation schemes.

To assess the laser capabilities to accelerate nanoparticles, the following experiment was performed. Vaterite nanocapsules were placed on a glass coverslip and positioned on an inverted microscope (Fig. 2(a)). A femtosecond laser (Menlo Systems, 100fs pulse, 1040nm wavelength, 100MHz repetition rate) was launched through a half-wave plate and polarizing beam splitter to control the light intensity illuminating the sample. To observe and track the release of the particles, a high-speed camera (Phantom v7.1, maximal frame rate 110kfps) was used to capture the dynamics. Two dichroic mirrors were used to isolate the light source and camera from the laser. A high-resolution scanning electron microscope (SEM) image of vaterite appears in Fig. 2(b), highlighting surface roughnesses of 2um-size spheres. While particles of those sizes were fabricated for better visualization purposes, smaller dimensions (below ~0.5µm) are favorable for drug delivery applications owing to cellular uptake aspects. Fig. 2(c) demonstrates a microscope image of the particles before and after the interaction with the laser, after which one item disappeared. The laser was focused on this exact spot. The minimum speed of particles was estimated from the depth of field of the objective (~0.8µm) and the frame rate. The particles were too fast for capturing their trajectory even at the highest frame rate of the camera. Consequently, only a lower bound on the velocities can be extracted from the experimental evidence. Since a particle disappeared from the field of view before the next image is captured, the minimal velocity is bounded from below by $\sim 22 \frac{cm}{s}$, which will be demonstrated hereinafter as a very underestimating.

To reveal whether the short pulses are responsible for the particle jump, a reference measurement has been performed. The femtosecond laser was replaced by a CW source (Cobolt Rumba 1064nm, 3W) and the experiment was repeated, keeping the average power of the CW and femtosecond sources the same. Since the jump of particles has not been observed under the CW illumination, the effect is unambiguously related to the high peak power, and a nonlinear time-dependent phenomenon is involved.

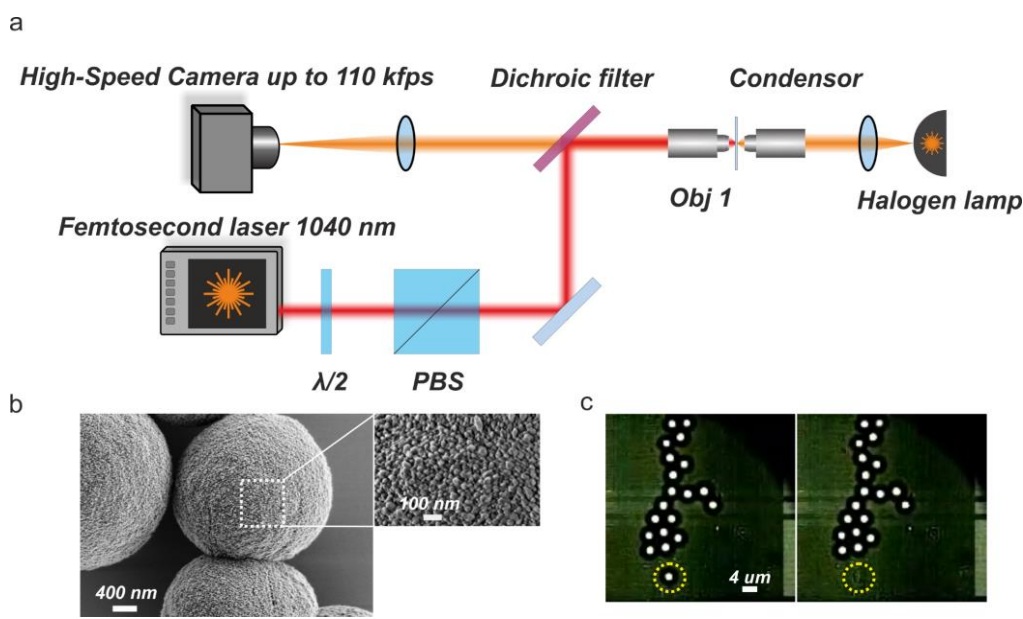

**Fig.2.** (a) Schematics of the experimental setup. (b) SEM images of vaterite particles. (c) Microscope images of vaterite particles on a glass substrate. Left - before interaction with the femtosecond laser. Right - after the interaction, demonstrating one particle missing.

## 3. Theoretical analysis

There are several parameters, which govern the interaction between the particle on the substrate and laser light. The first is the contact potential, which is responsible for sticking the particle to the substrate. In the air environment, Van der Waals forces are the key mechanism, which will dictate the laser power required for the jump. The second factor to reveal is the nature of the force, transferred from the laser beam to the particle. Hereinafter, we will provide a detailed analysis and discussion on the interaction nature. In particular, laser heating of vaterite particles and their fast thermal expansion will be proved to play a key role.

### 3.1 Van der Waals forces

Van der Waals forces bind a nanoparticle and a surface owing to intermolecular attraction. Since this interaction is extremely short-range, the laser-induced particle's jump has a threshold effect. In case of a short pulse illumination and low repetition rate, the particle will either remain on the surface or jump, depending on whether the energy in a pulse is sufficient. The threshold condition is given by:

$$\frac{mu_{min}^2}{2} - U_{vdW} = 0, \qquad (1)$$

where the minimal kinetic energy of a particle is equal to the van der Waals potential. Relying on the above discussion, the condition can be approximated as local i.e. after the particle leaves the surface, the Van der Waals potential can be neglected. In the case of a smooth sphere on a flat surface, van der Waals potential is given by [46]:

$$U_{vdW}(z) = -\frac{A_H R}{6z}, \qquad (2)$$

where $A_H$ is the Hamaker constant encapsulating the material properties, $R$ is the particle radius, $z > 0$ is the distance between the substrate and the particle surface. The calculation of the Hamaker constant appears in Supporting Information S1. Considering experiments done at normal conditions, as one reported here, a layer of water is always present in the contact area. Consequently, the distance between the particle and substrate surface can be approximated as $z = 0.3 nm$, corresponding to the size of a water molecule. This number can be also related to the particle's surface roughness (Fig. 2(b)). As a result, the force between a spherical R=2μm vaterite and a flat glass substrate is $F_{vdW}^{Vat} \approx -0.34 \mu N$. In this case, the gravity force $F_g \approx -1 pN$ can be neglected.

Following Eqs. 1 and 2, the threshold velocity is given by:

$$u_{min} = \frac{1}{2R}\sqrt{\frac{A_H}{\pi z_0 \rho}}, \qquad (3)$$

leading to $u_{min}^{Vat} \approx 5 \, cm/s$. Considering the scenario of a single pulse incidence, the physical meaning of this velocity is what the particle obtains whether the contact potential was zero. However, in case of a high repetition rate of 100MHz (given the field of view of the microscope and cm/sec-scale velocity estimate), the detached particle is hit by at least one more consecutive pulse, providing it with additional kinetic energy. Consequently, the experimental lower estimate and the theoretical prediction of the velocity, based on Eqs. 1 and 2, correlate well with each other.

Hereinafter, the momentum transfer mechanism will be discussed and the one, responsible for the effect, will be identified.

### 3.2 Light Momentum transfer mechanisms

Light-matter interaction phenomena involve many physical mechanisms, especially if nanostructures are involved. After performing experimental studies with several controlled parameters, four main probable mechanisms of light momentum transfer to the particle have been identified and summarized in Fig. 3, following reports of, e.g., [47]–[49].

The first one (Fig. 3(a)) is an optomechanical interaction [50], where light momentum is transferred to a particle via a coherent excitation of polarization in a material, leading to the emergence of macroscopic Lorentz force. Maxwell stress tensor is a tool, which is typically used for analysis. Optomechanically-driven particle jumps from substrate scenarios have been addressed theoretically [51]–[53] and experimentally [54], [55], considering intense laser pulses of at least several nanoseconds in duration. However, several special conditions have been met, which are not satisfied in our case. To reveal that this optomechanical mechanism cannot contribute to the observed effect, we have performed a numerical analysis, where optical forces were calculated in the time domain. We show that laser powers, used in the experiment, are insufficient to detach the particle from the surface. A detailed analysis of optical forces appears in Supporting Information S2.

The second probable mechanism is a Coulomb force, which emerges owing to an ultrafast charging of interacting materials due to a multiphoton ionization (Fig. 3(b)) [56], [57]. Charges are created on both the substrate and the particle, resulting in a repulsion. The third hypothesis is "shock waves" (Fig. 3(c)) [58]–[62]. Those are caused by a local ablation of a glass substrate, followed by a physical extraction of matter. In this case, the light should be focused on the substrate underneath the particle to cause a significant interaction. Both of those effects share similar underlying physical principles. The key mechanism is multi-photon absorption followed by ionization, which is a nonlinear process requiring high peak powers of an excitation source. Calcite has $E_g$=6 eV, which means that at least 5 photons are required for a non-linear multiphoton ionization caused by a 1040nm wavelength laser. However, the probability of such a process is negligible. The fluence in our experiment is ~0.15 J/cm$^2$, which is insufficient to make this process dominant. Also, the substrate was not damaged during the experiment, indicating that there is no matter extraction. This confirms that Coulomb repulsion and shock waves are not the main mechanisms of the laser-induced jump of particles.

The fourth mechanism is a thermal expansion (Fig. 3(d)) [63], [64]. Light absorption causes the material expansion and, owing to the substrate-induced broken symmetry, a fast elevation of the particle's center of mass occurs. Hereinafter, we will analyze this mechanism and demonstrate that it is indeed the one, responsible for the particle's jump.

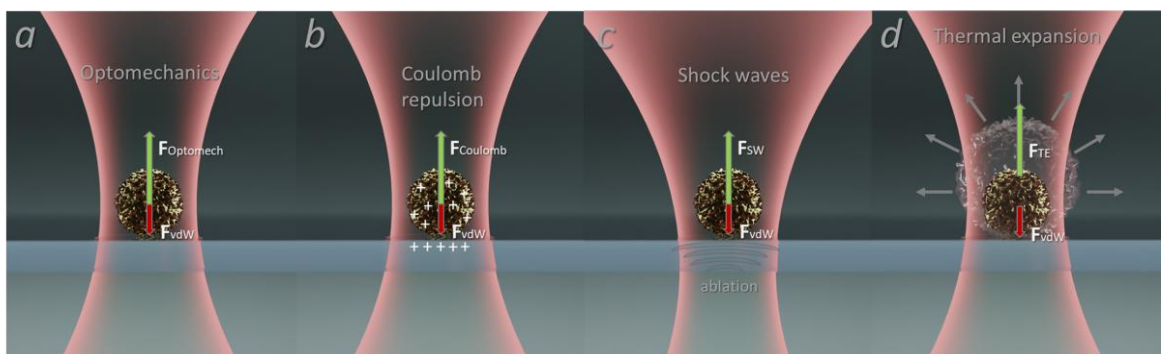

**Fig.3.** Schematics of four probable mechanisms, which can govern a laser-induced microparticle jump from the substrate: (a) Optomechanical force; (b) Coulomb force; (c) Material shock waves; (d) Optothermal expansion.

## 3.3 Momentum transfer by thermal expansion

Vaterite microcapsules have a relatively weak linear absorption, which was experimentally demonstrated in [65]–[67] without specification of the mechanism. The absorbed energy leads to a fast heating of the particle and, as a consequence, thermal expansion - reversible deformations of the lattice. Thermal effects can lead to the appearance of internal stresses in a microparticle, which manifest themselves as vibrations and expansions of the form factor [63]. Similar effects were observed considering other scenarios, e.g., the non-stationary displacements of the air-gold interface under the action of femtosecond laser radiation [68], [69]. The experimental technique was based on the probing beam deflection from the surface and demonstrated thermal expansion rise times of ~100 ps. The effect was attributed to the nonequilibrium diffusion and thermalization of photoexcited electrons.

In our arrangement, the fast deformation of a particle leads to a shift in its center of mass respectively to the surface, leading to the particle's jump. The stages of the process are shown in Fig. 4. Hereinafter, we will estimate the heating rate of a vaterite microparticle, considering its thermodynamic parameters and laser radiation intensity. Conditions, which are required for the particle's jump, will be estimated.

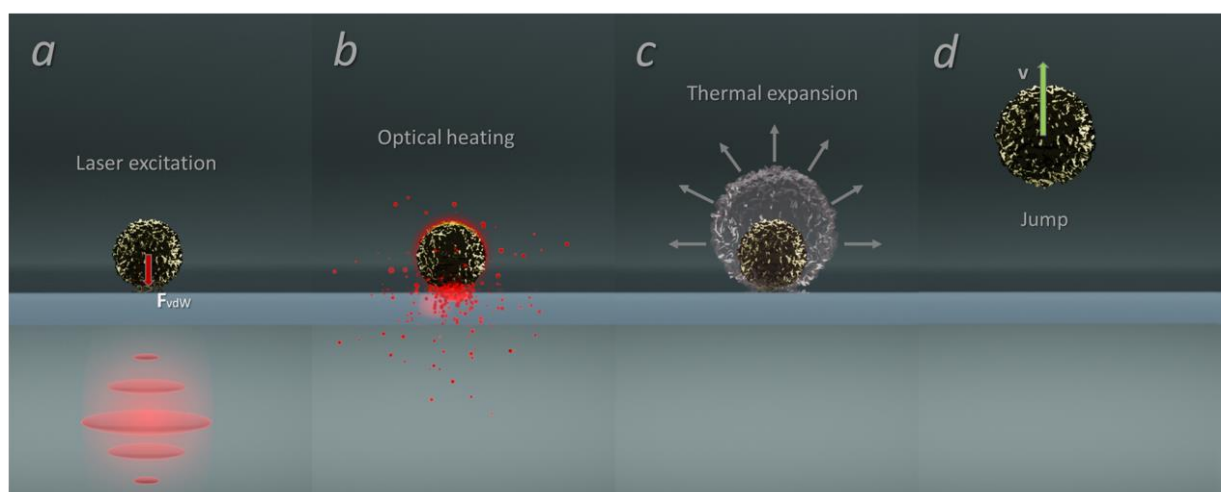

**Fig.4.** Stages of the optothermal jump - (a) laser excitation, (b) light absorption, (c) thermal expansion, and (d) jump. The processes (a) - (d) happen in parallel within a picosecond time scale.

In the case of short pulses, non-equilibrium phenomena may occur, nevertheless, the one-temperature model (OTM), being a simple approximation for the initial estimate [70]–[74], was found sufficient (see Supporting Information S3 and S4 for the details). For pulse durations lesser than 100 ps, the particle can be approximated as isolated since the heat removal by embedding air occurs on larger timescales. Consequently, the heat equation for the system is:

$$\rho C \frac{\partial T}{\partial t} = q(r,t), \quad (4)$$

where $T$ is the particle's temperature; $C$, $\rho$ — specific heat capacity, mass density; $q(r,t) = q_0 \exp\left[-\frac{2(t-t_0)^2}{\tau^2}\right]$ – is the volume density of the absorbed electromagnetic power, where $q_0$ is a constant, which depends on the material absorption and the laser intensity (Supporting Information S4 for the details). We also assume uniform heating of a nonresonant small particle. The absorbed power resembles the time envelope of the laser pulse (assumed to be Gaussian with the width $\tau = 100 fs$ ).

Eq. 4 has an analytical solution, which can be used to estimate the maximal temperature of the particle by taking the limit $(t \to \infty)$. The time here is longer than the pulse duration and transients inside the particle and shorter than the heat outflow to the air:

$$T_{MAX}^{OTM} = T_0 + \overbrace{\frac{\sqrt{\pi}\,\tau q_0}{2\sqrt{2}\rho C}\left[1 + \mathrm{erf}\left[2\sqrt{2}\right]\right]}^{const} = T_0 + \Delta T_{MAX}^{OTM}, \quad (5)$$

where $\mathrm{erf}\, x = \frac{2}{\sqrt{\pi}} \int_0^x e^{-\xi^2} d\xi$ is error function and $T_0$ is the ambient (room) temperature. The upper estimate for the center of mass velocity $u$ of the jumping the particle is:

$$u = \frac{dR}{dt} = \alpha \frac{dT}{dt} R_0, \quad (6)$$

where $\alpha$ - the material thermal expansion coefficient $[K^{-1}]$, and $R_0$ - particle's radius before interaction with the laser. Vaterite spherulite structure contains single nanocrystal subunits that are arranged as a bundle of fibers tied together at the center and spread out at the ends (so-called "heap of wheat" model [75]). While vaterite has a non-diagonal thermal expansion tensor with position-dependent components, we will use an experimentally verified isotropic averaged value, corresponding to calcite $\alpha_{zz}^C = 22.6 \cdot 10^{-6}\ K^{-1}$ [76]–[78], as a mere approximation. The estimated velocity acquired by a particle upon repulsion from the substrate as a result of thermal expansion is then given by:

$$u = \frac{\alpha_{zz}^{C} \Delta T_{MAX}^{OTM} R_0}{\tau}. \quad (7)$$

The detachment of a particle from the surface is of a threshold nature. If the linear momentum acquired by the particle as a result of the thermal expansion is sufficient to overcome the Van der Waals potential energy, then the particle jumps from the substrate with a nonzero velocity, otherwise, it remains bound:

$$v_z = \begin{cases} \dfrac{\alpha_{zz}^{C} \Delta T_{MAX}^{OTM} R_0}{\tau}, & u \geq u_{\min} \\ 0, & u < u_{\min} \end{cases}. \quad (8)$$

Eq. 8 allows estimating the conditions required for the particle jump. Fig. 5 demonstrates the velocity as a function of the pulse duration for several powers. Introducing the parameters of the experiment, we found that the velocities can be as high as tens of m/sec. This result does not contradict the initial experiment, as it solely provided the lower bound of 22cm/sec, which is now found to be not tight. m/sec scale velocities suggest assessing the capabilities of particles to penetrate into a phantom sample, which will be done next.

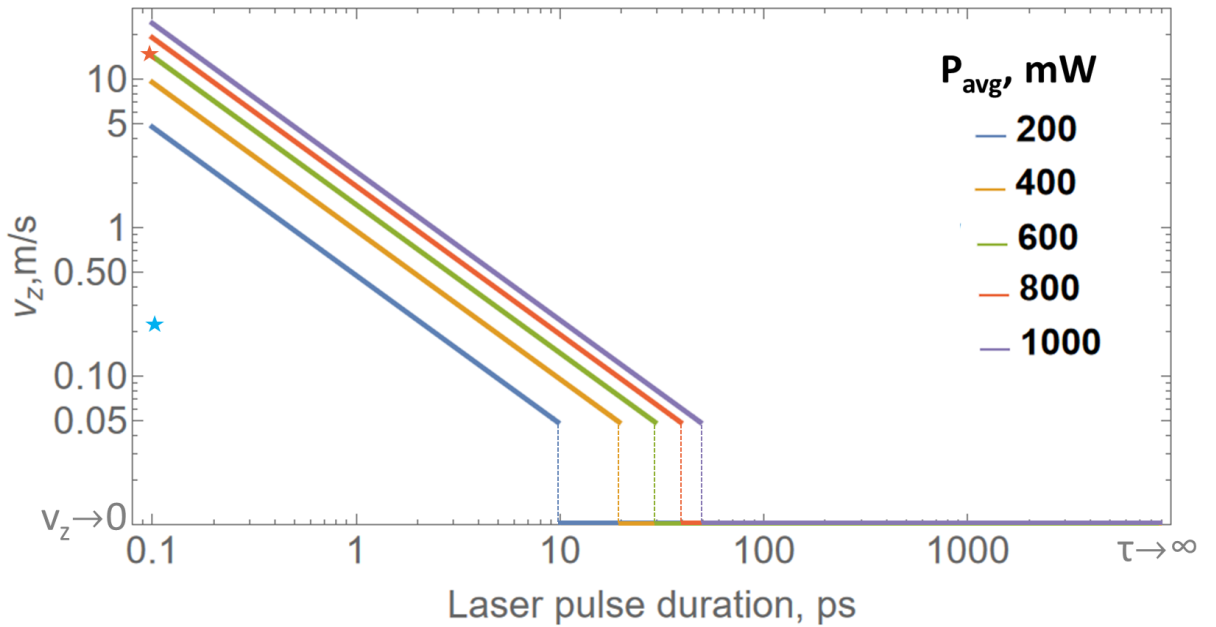

**Fig.5.** Particle's velocity as a function of laser pulse duration and average incident power (Eq. 8). Blue star marker - v ≈ 22 cm/s experimentally obtained a lower bound estimate (section 2). Red star marker - v ≈ 15 m/s a tighter bound assessed by a ballistic experiment (section 4).

## 4. Temperature Measurements

In order to ascertain the temperature induced in the particles upon illumination with femtosecond pulses we used fluorescence lifetime-based thermometry, e.g.,[79], [80]. The time-Correlated Single Photon Counting (TCSPC) method for measuring lifetimes has been used. A super-continuum laser (YSL SC-PRO, 300 ps repetition rate, 450-530 nm filter, 3mW average power to ensure not to induce an additional heating) was used as an excitation source of Rhodamine B. Vaterite particles were loaded with this dye by using a diffusion technique (e.g., [44]). The femtosecond laser was also launched on the sample through a beam splitter, and its intensity was kept as a variable. Intensities were kept below the threshold, required for the jump (Fig. 6(a)). The back-scattered fluorescence was filtered and assessed with the photon counter. Reference measurements were done by varying the temperature of the sample with a thermo-electric cooler. Ten different particles were measured to collect statistics and mitigate possible fluctuations.

Fig. 6(b) is the lifetime versus the environmental temperature. Red dots correspond to the reference measurement - the sample was heated to a temperature (lower x-axis on the plot) and the lifetime was retrieved. The next step is to illuminate the particles with the femtosecond laser. The averaged power is the upper x-axis on the plot. The blue solid curve is the trendline. The graph allows for relating the laser power to the particle's temperature by assessing x-axes vs each other. Since this temperature measurement corresponds to time averages and steady-state conditions, the process can be described with the heat diffusion equation, where the source is a CW illumination. The steady-state temperature is given by:

$$T_S = \frac{P_{avg} Q_{abs}}{4\pi R k_{Air}} + T_0, \tag{9}$$

where $Q_{abs} = \frac{P_{abs}}{P_{inc}} = 3.67 \cdot 10^{-4}$ is the absorptivity of the particle's material. Thus, for 50 mW laser power used in the experiment, the particle is heated to a temperature of about 60°C. This number is verified by the experiment (50 mW and 60°C are on top of each other in Fig 6(b)). Additional derivation can be found in Supporting Information S3 and S4.

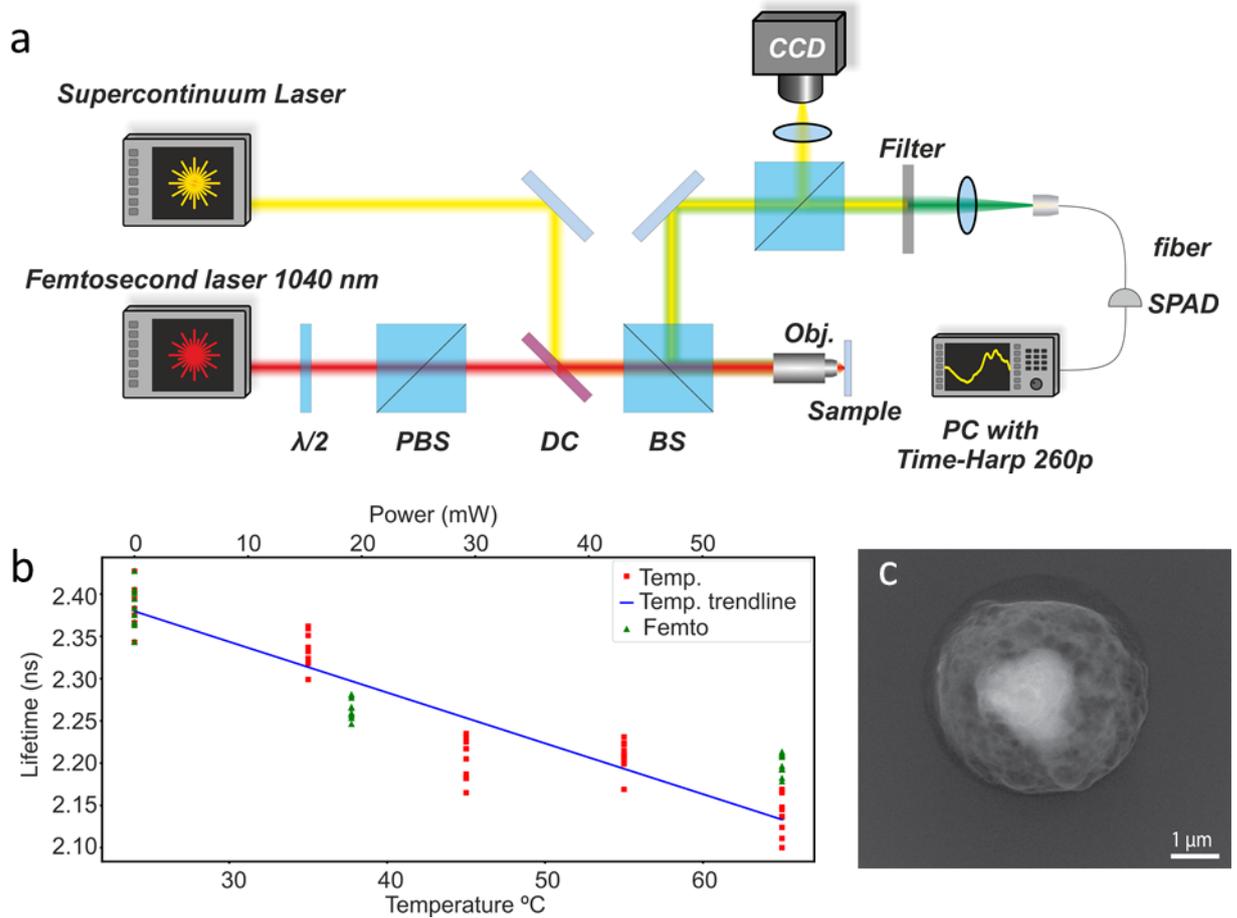

**Fig.6.** (a) Schematics of a setup for particle temperature measurements with fluorescent lifetime. (b) Lifetime-based particle thermometry with Rhodamine B, infused into vaterite. Red dots - reference lifetime and a function of the environmental temperature (lower x-axis). Green dots - lifetime as the function of an averaged laser poser (upper x-axis). Blue solid line - trendline. Each x-axis point data was measured for 10 particles. (c) SEM image of a 2um vaterite, stuck in SU8 polymer layer after the optothermal jump.

5. **Ballistic Experiment**

For a more accurate assessment of the particle's velocity and with an outlook on needle-free injection applications, the particle was jumped against a ballistic target. Using a setup similar to that in Fig. 2, we jumped Rhodamine-B-infused vaterite particles from a glass substrate upwards into a SU8 layer. The layer was positioned several millimeters above the sample. We then located the jumped particles in the layer using a confocal fluorescent microscope, marked their location, and then imaged the sample with SEM. Residuals of the particle with the SU8 layer can be seen in Fig. 6 (c). By assessing the depth of the crater within the polymer, the penetration depth of ~400-600nm can be deduced. To relate the results to a theoretical model, we used the Poncelet model [81], [82], and the maximum penetration depth is given by:

$$z_{max} = \frac{2}{3} \frac{\rho_s}{\rho_f} \frac{d}{C_D} \ln\left[\frac{\rho_f C_D u_0^2}{2\gamma} + 1\right] \qquad (10)$$

where $\gamma \sim 1 [MPa]$ - constant yield resistance (strength resistance); $C_D \approx \frac{24}{Re}$ - drag coefficient; $u_0$ - the impact velocity; $d = 4 [\mu m]$ - particle diameter; $\rho_{s/f} = 2500/1190 [kg/m^3]$ - density of particle/target; $\eta_0 \sim 1 [mPa \cdot s]$ - dynamic viscosity; $Re = \frac{\rho_f u_0 d}{\eta_0}$ - Reynolds number. Given the penetration depth and the rest of the parameters, the particle's velocity can be deduced. Substituting $z_{max}$ = 400–600 nm, as obtained in the experiment (Fig. 6(c)), an initial velocity is estimated as $u_0$ = 15–20 m/s. This result appears as a red dot in Fig. 5 and corresponds well with the rest of the data, thus verifying that the key mechanism beyond the observed effect is indeed optothermal.

## 6. Conclusion

Light-driven tools can provide a significant contribution to the personalized medicine paradigm by granting drug delivery capsules with extra functions. Here we explored the possibility to realize a needle-free injection scheme by accelerating a particle towards a target. Vaterite nanoparticles being among the most promising nonorganic drug delivery platforms have been explored. Apart from granting the particle sufficient velocity to penetrate the target, it has to be detached from a surface. To overcome a short-range Van der Waals potential binding the capsule to the substrate without using excessive laser powers, femtosecond pulses have been used. After the observation of the particle jump effect, a physical model for the process has been proposed and analyzed. Optomechanical, Coulomb, and Shock Wave mechanisms were eliminated after the detailed analyses of the experimental data. The mechanism of fast picosecond-scale optothermal expansion has been proposed, analyzed, and verified experimentally with the aid of nanothermometry tools, based on fluorescent lifetime imaging. Finally, the penetration capabilities of the particles were assessed by bombarding SU8 polymer layers. The Poncelet model was used to estimate particle velocities, which were found in 15m/sec range, given 0.5W average power 100fsec 1040nm laser. Being relevant to biomedical applications, the demonstrated concept further strengthens the capabilities of optical tools in drug delivery applications.

**Acknowledgments**

MIPT team gratefully acknowledge the financial support from the Ministry of Science and Higher Education of the Russian Federation (Agreement No. № 075-15-2022-1150). TAU team was supported by ERC StG"InMotion" (802279). There is no joint funding between the collaborating teams. The project was started in 2021. V.B. acknowledges the support of the Latvian Council of Science, project: PHOTON, No. lzp-2022/1-0579.

# SUPPORTING INFORMATION

## S1. Calculating Hamaker constants

The Hamaker constant can be calculated using the Lifshitz approach [1]:

$$A_H \cong \frac{3}{4}k_B T \left(\frac{\varepsilon_1(0)-\varepsilon_3(0)}{\varepsilon_1(0)+\varepsilon_3(0)}\right)\left(\frac{\varepsilon_2(0)-\varepsilon_3(0)}{\varepsilon_2(0)+\varepsilon_3(0)}\right) + \frac{3h}{4\pi}\int_{\nu_1}^{\infty}\left(\frac{\varepsilon_1(i\nu)-\varepsilon_3(i\nu)}{\varepsilon_1(i\nu)+\varepsilon_3(i\nu)}\right)\left(\frac{\varepsilon_2(i\nu)-\varepsilon_3(i\nu)}{\varepsilon_2(i\nu)+\varepsilon_3(i\nu)}\right)d\nu$$

(S1.1)

where $\nu$ - frequency of the electromagnetic field; $k_B$ - Boltzmann's constant; $T$ - temperature in Kelvin; $\varepsilon_k(0), k=1,2,3$ - static dielectric constant of media; $\varepsilon_k(i\nu), k=1,2,3$ - dynamic dielectric constant of media (1 - particle; 2 - substrate; 3 - environment); $\nu_1 = 2\pi k_B T/h = 3.9\times10^{13}\,Hz$ at $25°C$.

Expression (S1.1) takes into account both the entropy ($\nu=0$) and dispersion ($\nu>0$) contributions, thus dielectric properties of the all three materials have to be known. Moreover, we can use the approximations of Ninham and Parsegian [2] and Hough and White [3] to calculate the constant. If the dielectric medium has one strong absorption peak at a certain frequency $\nu_e$ (the average ionization frequency of the material, most often in the ultraviolet region, typical value $\nu_e \approx 3\cdot10^{15}\,Hz$), the dielectric constant can be approximated by:

$$\varepsilon(i\nu) = 1 + \frac{n^2-1}{1+\nu^2/\nu_e^2}$$

(S1.2)

where n is the refractive index. Considering materials with similar dispersion (e.g. transparent dielectrics), the undelayed Hamaker constant (Eq. S1.1) is approximated with:

$$A_H \approx \frac{3}{2}k_B T \left(\frac{\varepsilon_1(0)-\varepsilon_3(0)}{\varepsilon_1(0)+\varepsilon_3(0)}\right)\left(\frac{\varepsilon_2(0)-\varepsilon_3(0)}{\varepsilon_2(0)+\varepsilon_3(0)}\right) + \frac{3h\nu_e}{8\sqrt{2}}\frac{(n_1^2-n_3^2)(n_2^2-n_3^2)}{\sqrt{n_1^2+n_3^2}\sqrt{n_2^2+n_3^2}\left(\sqrt{n_1^2+n_3^2}+\sqrt{n_2^2+n_3^2}\right)}$$

(S1.3)

So, the main contribution to the Hamaker constant comes from the visible and UV spectral ranges.

The table shows the parameters for calculating the constant [4]:

|  | $\varepsilon(0)$ | $n$ |
|---|---|---|
| medium 1 vaterite (CaCO$_3$) | 8.2 | 1.6 |
| medium 2 (glass) | 3.82 | 1.46 |
| medium 3 (air) | 1 | 1 |

For smooth vaterite particles, the Hamaker constant is $A_H^{Vat} \approx 9\cdot10^{-20}\,J$.

## S2. Optomechanical interactions with a femtosecond laser pulse

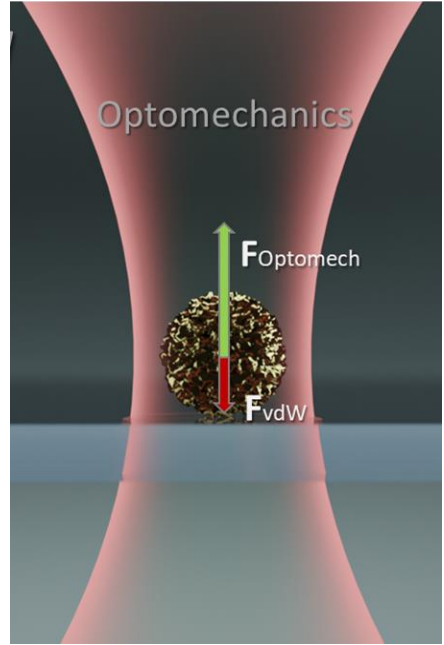

**Fig.S2.1** Sketch of the optomechanical interaction between a particle on a substrate and a laser pulse.

The temporal dynamics of optical forces can be assessed with the formalism of time-dependent Maxwell stress tensor. The electromagnetic optical force density is given by [5]:

$$f_i = \partial_j T_{ij} - \partial_t g_i \tag{S2.1}$$

The instantaneous force density $f_i$ consists of two terms, both related to the transfer of momentum from the electromagnetic field to the particle. The first term corresponds to the time-dependent Maxwell stress tensor:

$$T_{ij} = E_i D_j + H_i B_j - \frac{1}{2}\delta_{ij}\left(E_k D_k + H_k B_k\right) \tag{S2.2}$$

The second term in Eq. S2.1 is the momentum carried by an electromagnetic field within the volume V. It originates from the dynamic terms in Maxwell's curl equations, containing the time derivatives:

$$g_i = \varepsilon_{ijk} D_j B_k \tag{S2.3}$$

Here $\varepsilon_{ijk}$ is a completely antisymmetric Levi-Civita pseudotensor; $E_i, D_i, H_i, B_i$ - Cartesian components of electric and magnetic fields and inductions.

To calculate the total optical force acting on a structure, Eq. S2.1 has to be integrated over the particle's volume. Applying Gauss theorem, we obtain [6]:

$$\mathbf{F}_{pulse}(t) = \frac{d\mathbf{p}}{dt} = \int_{\partial V} \vec{T}(\mathbf{r},t) \cdot \mathbf{n}(\mathbf{r}) ds - \frac{d\mathbf{p}_{field}}{dt}, \qquad \mathbf{p}_{field} = \frac{1}{c^2} \int_V [\mathbf{E} \times \mathbf{H}] dV \qquad (S2.4)$$

Hereinafter, the momentum transfer to a particle illuminated with a single $\tau_0 = 100\,fs$ pulse centered at $\lambda_0 = 1040\,nm$ will be considered. Since the glass slide reflectivity is ~ 4%, the substrate can be neglected, and optical forces acting on an isolated particle (without the substrate) will be calculated. In the numerical setup, the particle is placed at the center of the Gaussian beam (radius $w_0 = 2\,\mu m$). In this case, only the radiation pressure will act on the particle, while the gradient force vanishes.

It should be noted that vaterite particles have anisotropic dielectric tensors. However, in [7], the internal structure of a spherulite was shown to demonstrate a moderately low impact on scattering patterns, i.e. the orientation of the particle with respect to the incident beam can be neglected for the estimation.

The dynamics of the momentum transfer to a particle can be calculated based on eq. (S2.4) as:

$$\mathbf{p}(t) = \int_0^t \mathbf{F}_{pulse}(t) dt \qquad (S2.5)$$

Figure S2.2 demonstrates the time-dependent accumulated momentum for different average laser powers (in legends) and shows the result of the interaction of a short laser pulse with vaterite microparticles.

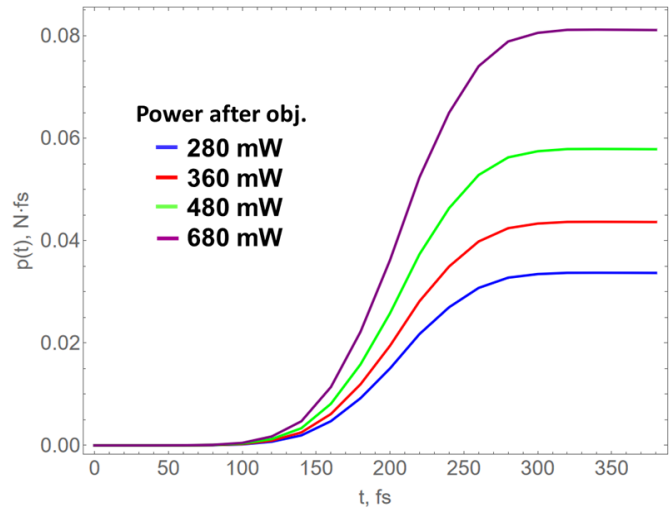

**Fig.S2.2.** Dynamics of the mechanical moment transfer from a short laser pulse for vaterite particle ($n = 1.6$). Laser parameters: $\tau_0 = 100\,fs$ - pulse duration; $\lambda_0 = 1040\,nm$ - the laser wavelength; $w_0 = 2\,\mu m$ - the radius of the beam waist. Particle radius $R_0 = 2\,\mu m$.

Fig. S2.3 demonstrates the maximal optomechanical momentum as a function of the laser average power. The horizontal dashed line is a threshold momentum required for the jump (equivalent to the van der Waals force attracting particles to the substrate) (see eq. 3 in the main text). Comparing the threshold momentum with the results suggests that particles cannot be detached from the surface if the primary physical mechanism is pure optomechanics.

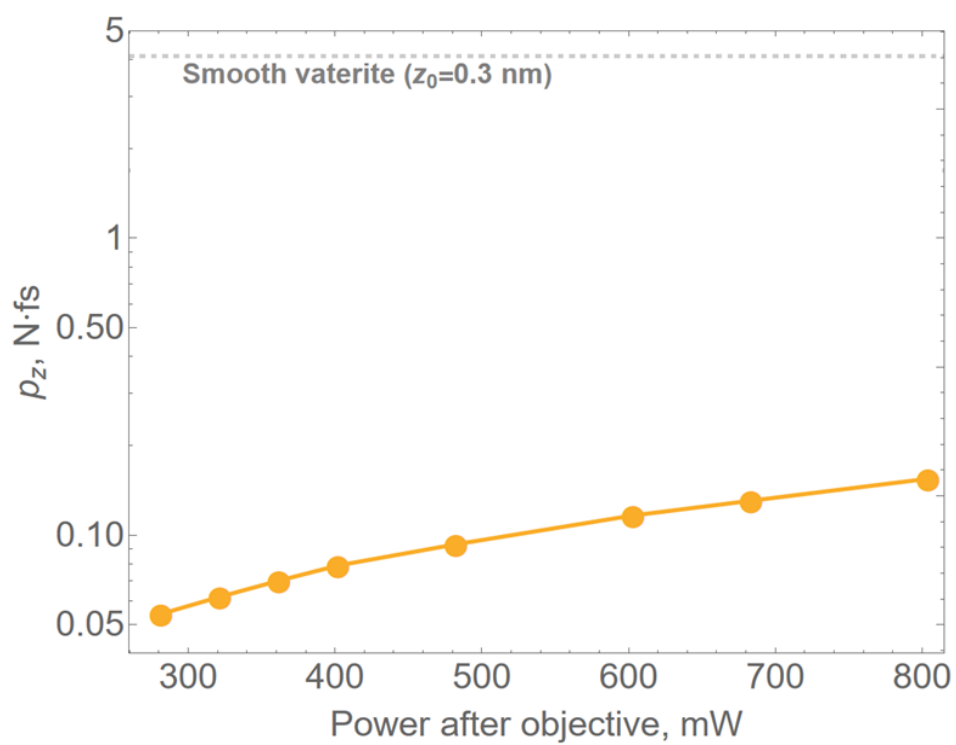

**Fig. S2.3.** Momentum, transferred from a femtosecond laser pulse to a vaterite particle, as a function of the laser power. The dotted line indicates the threshold momentum.

## S3. Vaterite Absorption

Vaterite microcapsules have relatively weak linear absorption [8]–[10]. The data was obtained by exploring spherical particles under CW illumination.

Table S3.1 – Experimental parameters from [9] and [10].

|  | **Parkin et. al.** [9] | **Dholakia et. al.** [10] |
|---|---|---|
| **Particle radius, μm** | 1.6 | 0.423 |
| **Wavelength, nm** | 1064 | 532 |
| **Beam waist, μm** | 0.5 | 0.26 |
| **Host media** | MetOH | $D_2O$ |
| **Temperature (at 1 W), °C** | 66 | 4.2 |
| **Thermal conductivity, $Wm^{-1}K^{-1}$** | 0.202 | 0.61 |

The imaginary part of the refractive index (n'') is responsible for heating. $n''_{1040} \sim 10^{-6}$, corresponding to calcite, was used for the estimations [11, 12]. Numerical analysis corresponding to parameters in Table S3.1 is summarized graphically in Fig. S3.1.

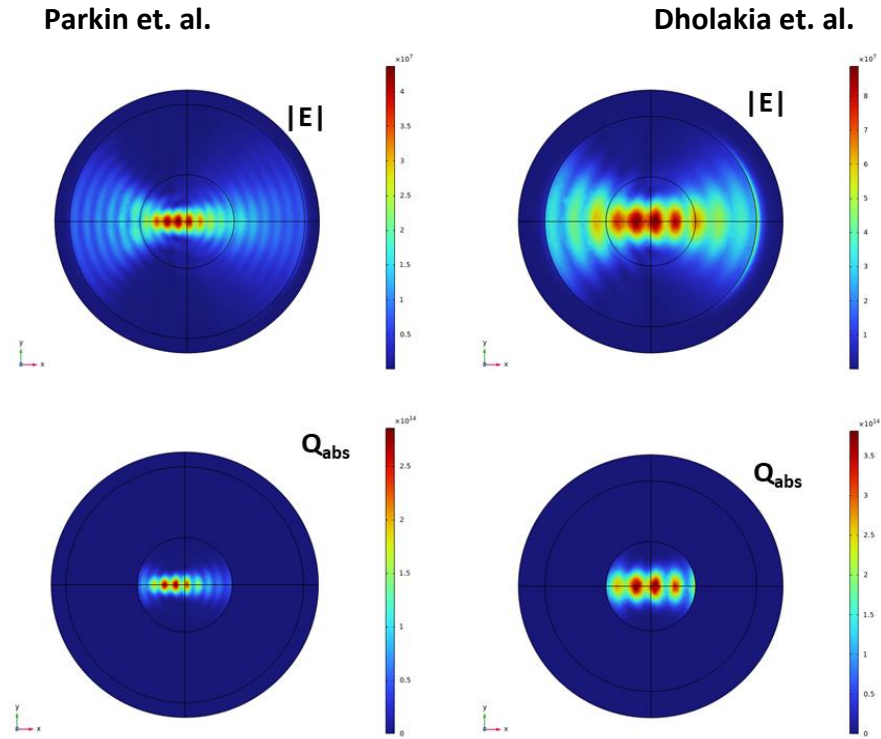

**Fig.S3.1.** Electromagnetic interactions with microparticles. Details are in Table S3.1 and legends.

As a result of simulations, the following values were obtained: $n''^{(1064)}_{vat} = 8.5 \cdot 10^{-6}$ and $n''^{(532)}_{vat} = 9.7 \cdot 10^{-8}$. These values were used in the further analysis $n''^{(1040)}_{vat} \approx n''^{(1064)}_{vat} = 8.5 \cdot 10^{-6}$.

The particle's temperature was also calculated analytically.

The model is based on a simple system of stationary heat diffusion equations in spherical coordinates:

$$\begin{cases} \dfrac{1}{r^2}\dfrac{d}{dr}\left(r^2 \dfrac{dT_1}{dr}\right) = -\dfrac{q_0}{k_1} & \text{inside NP} \\ \dfrac{1}{r^2}\dfrac{d}{dr}\left(r^2 \dfrac{dT_2}{dr}\right) = 0 & \text{outside NP} \end{cases} \quad (S3.1)$$

with boundary conditions

$$\begin{cases} T_1\big|_{r=R} = T_2\big|_{r=R} \\ T_1\big|_{r=0} < \infty \\ T_2\big|_{r\to\infty} = T_0 \\ -k_1 \dfrac{dT_1}{dr}\bigg|_{r=R} = -k_2 \dfrac{dT_2}{dr}\bigg|_{r=R} \end{cases} \quad (S3.2)$$

This system has the exact solution:

$$\begin{cases} T_1 = -\dfrac{q_0 r^2}{6k_1} + \dfrac{q_0 R^2}{3}\left(\dfrac{1}{k_2} + \dfrac{1}{2k_1}\right) + T_0 \\ T_2 = \dfrac{q_0 R^3}{3k_2}\dfrac{1}{r} + T_0 \end{cases} \quad (S3.3)$$

Then the temperature on the surface of a particle:

$$T_S(R) = \dfrac{q_0 R^2}{3k_2} + T_0 \text{ or } T_S(R) = \dfrac{P_{abs}}{4\pi k_2 R} + T_0 \quad (S3.4)$$

The absorbed power is given by:

$$P_{abs} = \frac{\omega}{2}\varepsilon_0 \operatorname{Im}\varepsilon \int_{V_{NP}} |\mathbf{E}(\mathbf{r})|^2 dV \qquad (S3.5).$$

The absorptivity of the particle's material is defined as the ratio of the power absorbed in the particle to the power of the pulse $Q_{abs} = \frac{P_{abs}}{P_{inc}}$. To simplify the calculations, we assume the energy being absorbed uniformly $P_{abs} = q_0 V$, leading to $q_0 = \frac{3 P_{abs}}{4\pi R^3}$. For CW laser $q_0 = \frac{3 P_{avg} Q_{abs}}{4\pi R^3}$.

The simulation results are shown in the graphs below (Fig. S3.2). The solid curve is the analytical formula verified by the full-wave analysis (blue solid line).

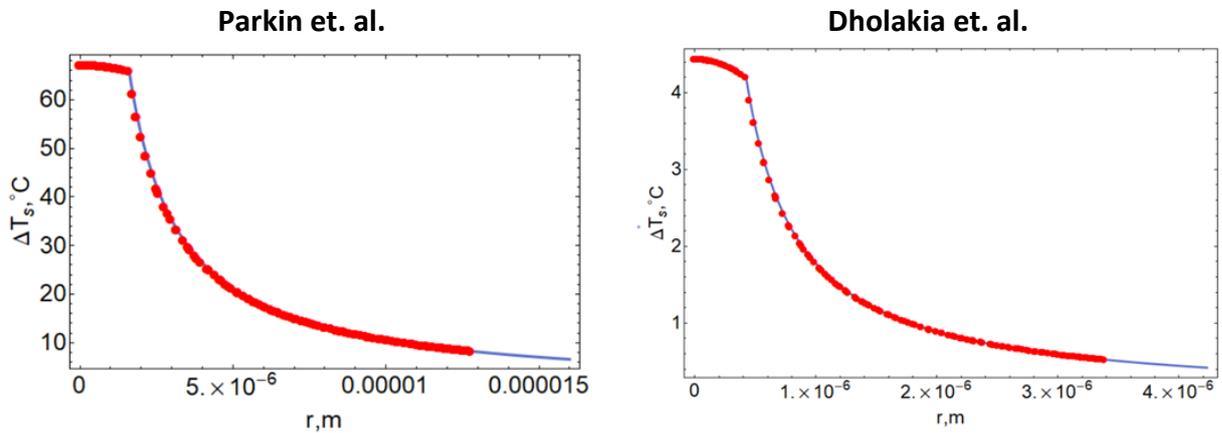

**Fig.S3.2.** Temperature profiles for 1W laser power.

Temperature profile of a heated vaterite particle calculated based on the parameters in Table S3.2 and fitted values of the imaginary part of the complex refractive index. Red dots - modeling in COMSOL; solid lines are the spatial distribution of temperature, calculated by formula S3.3.

Further, using the obtained values of the imaginary part of the refractive index, a similar numerical simulation was carried out for a vaterite particle with parameters taken from our experiment.

Table S3.2 – Model parameters

| Particle radius, µm | 2 | |
|---|---|---|
| Wavelength, nm | 1040 | |
| Beam waist, µm | 2 | |
| Host media | Air | H$_2$O |
| Thermal conductivity, Wm$^{-1}$K$^{-1}$ | 0.022 | 0.61 |

Numerical results demonstrating temperature profits at different conditions are summarized in Fig. S3.3.

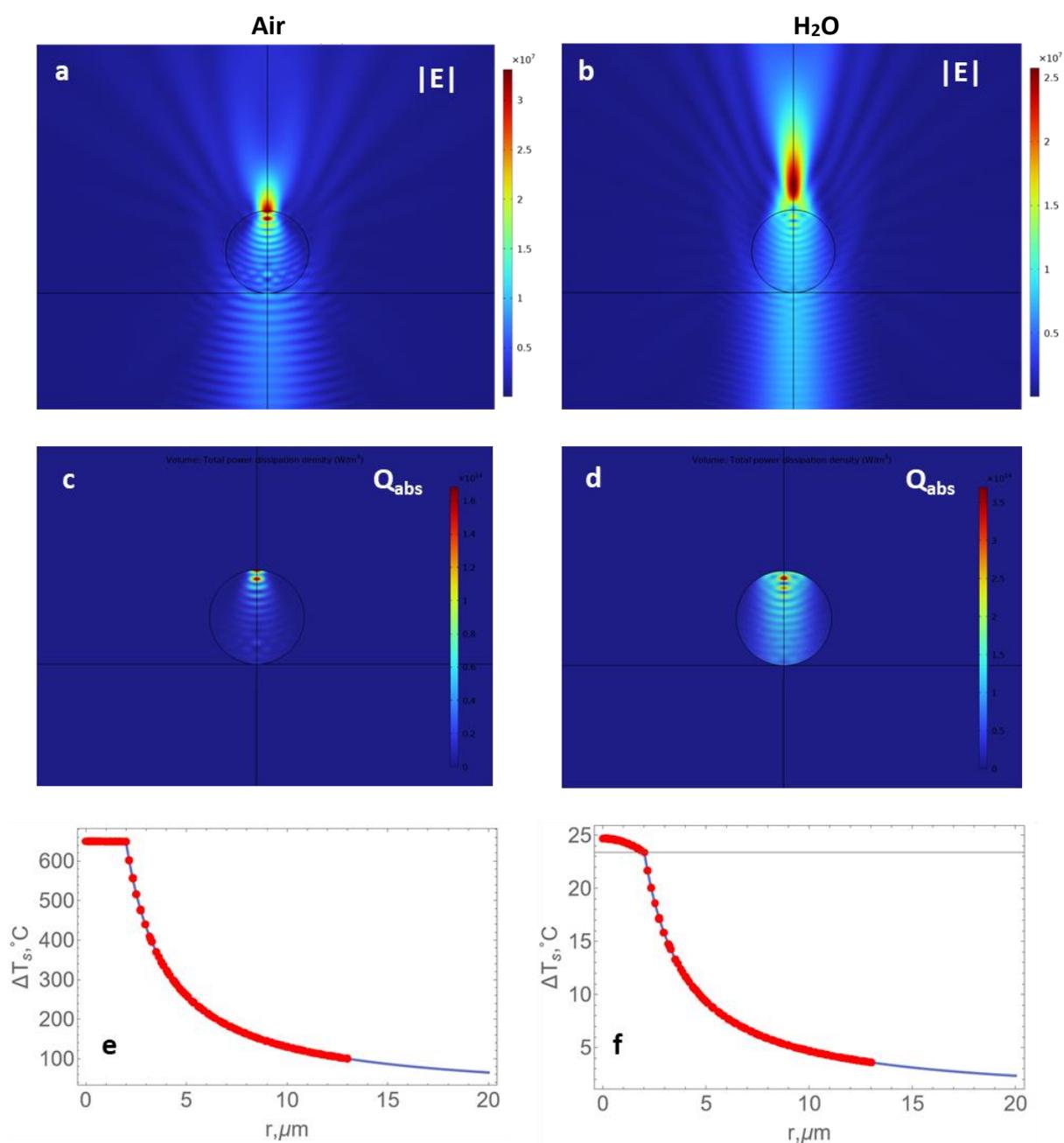

**Fig.S3.3.** Results of numerical simulation. (a, b) Spatial distribution of the electric field for a vaterite particle located on a glass substrate. The system is irradiated with a linearly polarized Gaussian beam. The beam propagates from below. (c, d) Spatial distribution of the volume energy density absorbed by the particle (lossless substrate). (e, f) Temperature profile of a heated vaterite particle irradiated with a 1 W CW laser. Red dots - COMSOL; the solid line is the values obtained from formula S3.3.

## S4. Pulsed laser heating of a particle.

In the case of short pulses, nonequilibrium phenomena govern the interaction. One temperature model is a mere approximation for the process [13]–[17]:

$$\begin{cases} \rho_l C_l \dfrac{\partial T_l}{\partial t} = k_l \dfrac{1}{r^2} \dfrac{\partial}{\partial r}\left(r^2 \dfrac{\partial T_l}{\partial r}\right) + q(r,t) \\ \rho_m C_m \dfrac{\partial T_m}{\partial t} = k_m \dfrac{1}{r^2} \dfrac{\partial}{\partial r}\left(r^2 \dfrac{\partial T_m}{\partial r}\right) \end{cases}, \quad (S4.1)$$

where $T$ is the system's temperature, while subindexes $m$ and $l$ stay for an embedding medium and the object's lattice, respectively. $C$, $\rho$, and $k$ are specific heat capacity, mass density, and heat conductivity. $q(r,t)$ is the absorbed electromagnetic power density in the volume of a particle.

Boundary conditions are:

$$\begin{cases} \left.\dfrac{\partial T_l}{\partial r}\right|_{r\to 0} = 0 \\ T_l(R,t) = T_m(R,t) \\ T_m(\infty,t) = T_0 \\ T_l(0,t) < \infty \\ \left.k_l \dfrac{\partial T_l}{\partial r}\right|_{r=R} = \left.k_m \dfrac{\partial T_m}{\partial r}\right|_{r=R} \end{cases}. \quad (S4.2)$$

Initial conditions:

$$\begin{cases} T_l(r,0) = T_0 \\ T_m(r,0) = T_0 \end{cases}. \quad (S4.3)$$

Thermodynamic parameters for describing vaterite in air are:

$$\begin{cases} k_l = 5.5[W/m\cdot K] & k_m = 0.022[W/m\cdot K] \\ \rho_l = 2500[kg/m^3] & \rho_m = 1.2[kg/m^3] \\ C_l = 830[J/kg\cdot K] & C_m = 1005[J/kg\cdot K] \end{cases}. \quad (S4.4)$$

The heat source resamples the form of the incident pulse:

$$q(r,t) = q_0 f(t) = q_0 \exp\left[-\frac{2(t-t_0)^2}{\tau^2}\right], \tag{S4.5}$$

where $q_0$ is a constant depending on the material absorption coefficient and the laser intensity. Taking into account the relationship between the power in a pulse and the average energy and the time parameters of laser $P_{inc}^{pulse} = \dfrac{P_{avg}}{f_{rep}\tau}$, we obtain:

$$q_0 = \frac{3Q_{abs} P_{avg}}{4\pi R^3 f_{rep} \tau}. \tag{S4.6}$$

Figures S4.1 a and b show the results of the electromagnetic simulations. It can be seen that the field is focused by the particle formingt he so-called photonic nanojet. This leads to the creation of «hot spots". To simplify the analysis, we will approximate the process with a uniform distribution

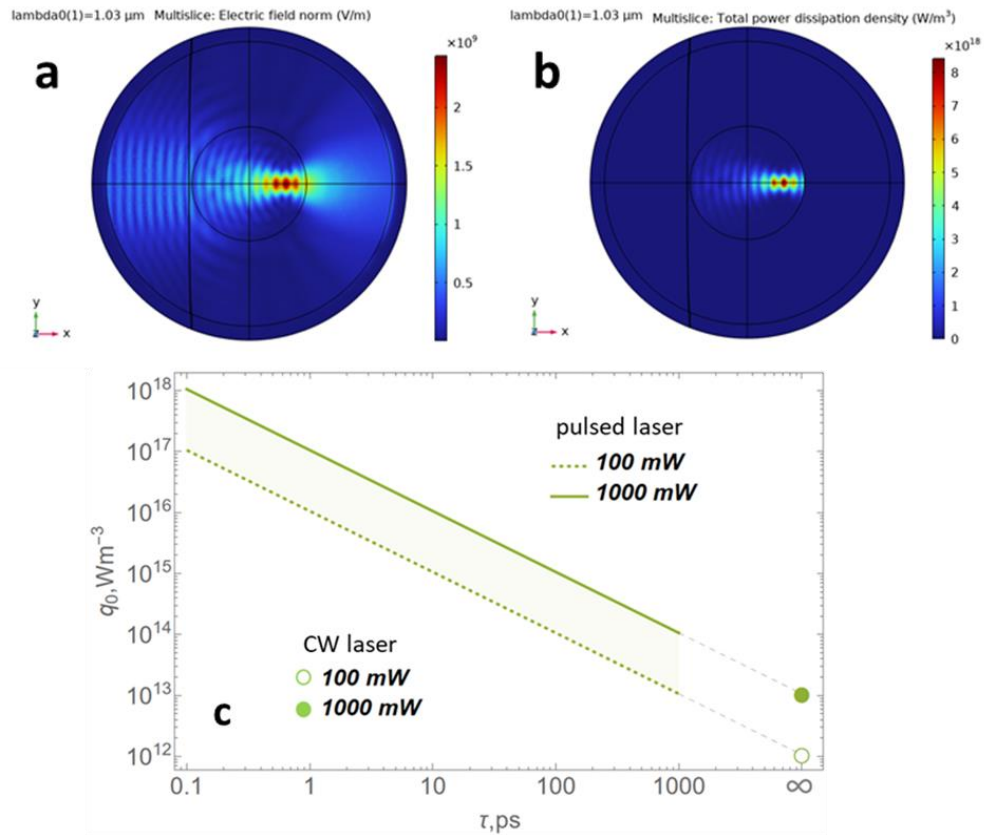

**Fig. S4.1.** (a) electric field norm; (b) volume density of the absorbed electromagnetic power. (c) The value of $q_0$ constant as a function of the pulse duration for the average laser powers, ranging from 100 to 1000 mW. The dots show the value of the $q_0$ constant for a CW laser.

In this case, you can get an analytical solution:

$$T^{OTM}(t) = T_0 + \frac{\sqrt{\frac{\pi}{2}}\tau\, Erf\left[2\sqrt{2}\right]q_0}{2\rho C} + \frac{\sqrt{\frac{\pi}{2}}\tau\, Erf\left[\frac{\sqrt{2}(t-t_0)}{\tau}\right]q_0}{2\rho C} \qquad (S4.7)$$

An analytical estimate of the temperature of a particle as a result of the absorption of a laser pulse of various intensities is shown in Figure S4.2. The calculations were performed for the parameters implemented in the experiment.

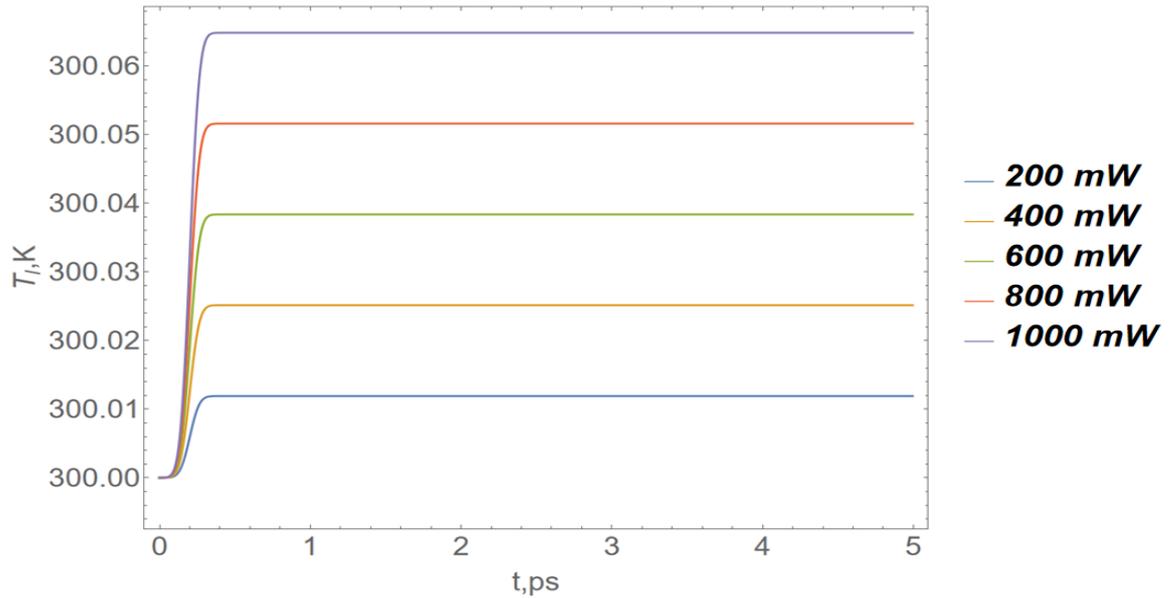

**Fig. S4.2.** The time dependent temperature of a particle when irradiated with a laser pulse of different energies, calculated by formula (S4.7). Modeling parameters $f_{rep} = 100\,MHz$, $w_0 = 2\,\mu m$, $R_{NP} = 2\,\mu m$, $T_0 = 300\,K$, $Q_{abs} = 3.67 \cdot 10^{-4}$, $\tau = 100\,fs$